# Imaging Dirac-Mass Disorder from Magnetic Dopant-Atoms in the Ferromagnetic Topological Insulator $Cr_x(Bi_{0.1}Sb_{0.9})_{2-x}Te_3$

Inhee Lee[1†], Chung Koo Kim[1†], Jinho Lee[1,2], S. J. L. Billinge[1,3], R. D. Zhong[1,4], J. A. Schneeloch[1,5], T. S. Liu[1,6], T. Valla[1], J. M. Tranquada[1], G. D. Gu[1], and J. C. Davis[1,7,8,9]

1. CMPMS Department, Brookhaven National Laboratory, Upton, NY 11973, USA.
2. Department of Physics and Astronomy, Seoul National University, Seoul 151-742, Korea.
3. Department of Applied Physics and Applied Mathematics, Columbia University, NY, NY 10027, USA.
4. Materials Science and Engineering Dept., Stony Brook University, Stony Brook, NY 11794, USA.
5. Department of Physics and Astronomy, Stony Brook University, Stony Brook, NY 11794, USA.
6. School of Chemical Engineering and Environment, North University of China, Shanxi 030051, China.
7. LASSP, Department of Physics, Cornell University, Ithaca, NY 14853, USA.
8. School of Physics and Astronomy, University of St. Andrews, Fife KY16 9SS, Scotland.
9. Kavli Institute at Cornell for Nanoscale Science, Cornell University, Ithaca, NY 14853, USA.
† These authors contributed equally to this work.

**To achieve and utilize the most exotic electronic phenomena predicted for the surface states of 3D topological insulators (TI), it is necessary to open a "Dirac-mass gap" in their spectrum by breaking time-reversal symmetry. Use of magnetic dopant atoms to generate a ferromagnetic state is the most widely used approach. But it is unknown how the spatial arrangements of the magnetic dopant atoms influence the Dirac-mass gap at the atomic scale or, conversely, whether the ferromagnetic interactions between dopant atoms are influenced by the topological surface states. Here we image the locations of the magnetic (Cr) dopant atoms in the ferromagnetic TI $Cr_{0.08}(Bi_{0.1}Sb_{0.9})_{1.92}Te_3$. Simultaneous visualization of the Dirac-mass gap $\Delta(r)$ reveals its intense disorder, which we demonstrate directly is related to fluctuations in n($r$), the Cr atom areal density in the termination layer. We find the relationship of surface-state Fermi wavevectors to the anisotropic structure of $\Delta(r)$ consistent with predictions for surface ferromagnetism mediated by those states. Moreover, despite the intense Dirac-mass disorder, the anticipated relationship $\Delta(r) \propto n(r)$ is confirmed throughout, and exhibits an electron-dopant interaction energy $J^*$=145 meV·nm$^2$. These observations reveal how magnetic dopant atoms actually generate the TI mass gap locally and that, to achieve the novel physics expected of time-reversal-symmetry breaking TI materials, control of the resulting Dirac-mass gap disorder will be essential.**

That the surface states of three-dimensional TI's exhibit a 'massless' Dirac spectrum $H(\mathbf{k}) = \hbar v \mathbf{k} \cdot \boldsymbol{\sigma}$ with spin-momentum locking and protected by time-reversal symmetry, is now firmly established. Opening a gap in this spectrum is key to the realization of several extraordinary new types of electronic phenomena. The prevalent approach to opening this 'Dirac-mass gap' is to dope the materials with magnetic atoms (*1-6*). A plethora of new physics is then predicted, including $\sigma_{xy} = \pm e^2/h$ quantum anomalous Hall effects (QAHE) (*7,8*), topological surface state magneto-electric effects (*9–12*), related magneto-optical Kerr and Faraday rotations (*10,13,14*), axionic-like electrodynamics (*15,16*), and even **E**-field induced magnetic monopoles (*17,18*). As yet, excepting the QAHE, none of these phenomena have been detected.

Interactions between the surface electrons and the magnetic dopant atoms at random surface locations $\mathbf{R}_i$ can be represented theoretically by a Hamiltonian of the type $H_{DA} = -J^* \sum \mathbf{S}_i \cdot \mathbf{s} \delta(\mathbf{r} - \mathbf{R}_i)$. Here $\mathbf{S}_i$ (**s**) is the spin of each dopant (surface-state carrier) measured in units of $\hbar$, and $J^*$ is their exchange-interaction energy scale. In the simple case of a homogenous ferromagnetic state with magnetization parallel to the surface normal $\hat{z}$, the Hamiltonian becomes $H = -J^* n_0 S_z \sigma_3/2$ where $n_0$ is the average two-dimensional dopant-atom density and $S_z$ the magnitude of the z-component of the dopant-spin. Such interactions should open a Zeeman-like energy gap of magnitude $\Delta = J^* M_z/2\mu_B (\equiv mv^2)$ where $M_z = n_0 S_Z \mu_B$ is the homogeneous $\hat{z}$-aligned magnetization, *m* is the Dirac-mass, and *v* the Fermi velocity. The resulting surface state dispersion is given by $E_{\pm}(k) = E_D \pm \sqrt{(\hbar v)^2 k^2 + \Delta^2}$ relative to the surface-state Fermi energy $E_F$, with $\Delta$ the Dirac-mass gap. Angle resolved photoemission studies provide good evidence that high densities of magnetic dopant atoms generate a ferromagnetic state and open such energy gaps in TI materials (*19,20*). Nevertheless, theoretical studies of dopant effects (*1-6*) have raised several fundamental issues about the atomic-scale phenomenology of ferromagnetic TIs that can only be resolved by direct electronic structure visualization experiments. First, what effect (if any) does the random distribution of dopant atoms have on the formation and homogeneity of the ferromagnetic state? Second, and perhaps most importantly, what are the consequences of any nanoscale disorder in the ferromagnetism for spatial arrangements of Dirac-mass gap? Finally, if such Dirac-mass disorder existed, how would it influence the all-important transport characteristics of the surface states? A detailed atomic-scale understanding of the actual physical arrangements of ferromagnetic TI surface states in the presence of magnetic dopant atoms is required to understand these issues.

Theoretical models (*1-6*) for the surface physics of a doped TI hypothesize that magnetic interactions between pairs of dopant atoms are mediated by the topological surface states (*SI* section 1). Elementary models estimate (*1*) the effective *z*-component of magnetic field at dopant site *i* as

$$B_{z,i}(\boldsymbol{r}) = \sum_j \Phi_Z(\boldsymbol{r}-\boldsymbol{r}_j) S_{z,j} \tag{1A}$$

where $\boldsymbol{r}_j$ are the dopant-atom locations and

$$\Phi_Z(\boldsymbol{r}) = -J^{*2}(F_+(k_F r) + (F_-(k_F r))/\hbar v r^3 \tag{1B}$$

Here $k_F$ is the Fermi wavevector of the topological surface states, $F_{+(-)}(x_F) = \int_0^{x_F(x_c)} \frac{x\,dx}{2\pi} \int_{x_F}^{x_c} \frac{x'dx'}{2\pi} \frac{1}{+(-)x-x'} [J_0(x)J_0(x') - (+)J_1(x)J_1(x')]$ where $J_n(x)$ is the Bessel function, and $x_c = k_c r$ effectuates a high cut-off momentum $k_c$. In such cases, the distribution of *z*-axis magnetization $M_Z(\boldsymbol{r})$ should become heterogeneous due to random fluctuations in spatial density $n(\boldsymbol{r})$ of dopant atoms. The most important consequences of such spatial variations in $M_z(\boldsymbol{r})$, whatever their microscopic origin, include a heterogeneous Dirac-mass gap

$$\Delta(\boldsymbol{r}) = -J^* M_z(\boldsymbol{r})/2\mu_B \tag{2}$$

(*SI* section 1) and thus a spatially disordered dispersion of the surface states $E_\pm(\boldsymbol{r}) = E_D(\boldsymbol{r}) \pm \sqrt{(\hbar v)^2 k^2 + \Delta(\boldsymbol{r})^2}$. Although excellent progress has been achieved in visualization of electronic structure in ferromagnetic TI compounds (*21,22,23*), the atomic-scale effects of magnetic dopant atoms on the Dirac-mass gap and on its contingent physics have not yet been determined.

To explore these issues, we use spectroscopic imaging scanning tunneling microscopy (*24*) (SI-STM) to study $Cr_{0.08}(Bi_{0.1}Sb_{0.9})_{1.92}Te_3$ (CBST) single crystals. These materials are chosen because they are indeed ferromagnetic (*25,26*), they exhibit topological surface states with a Dirac point $E_D$ near the Fermi energy $E_F$ (*27*), and exhibit the QAHE (*28,29*,30). Our SI-STM experiments are carried out in cryogenic ultra-high vacuum at T=4.5K, well below the measured $T_C \sim 18K$ bulk ferromagnetic phase transition in these samples. The technique consists of measuring the differential tunneling conductance $dI/dV(r, E = eV) \equiv g(r, E)$ as a function of both location $r$ and electron energy $E$; it is unique in capability to access simultaneously the $r$-space and $k$-space electronic structure for states both above and below the Fermi energy (24). In Fig. 1A we show a typical topographic image T($r$) of the Te termination surface of our CBST samples. Figure 1B is a schematic of the relevant unit cell and identifies the Cr atom substitutional site (pink) in the Bi-Sb layer just beneath the Te surface; we see that it occurs in the center of a triangle of surface Te atoms. This allows the location of each Cr dopant atom adjacent to the termination layer, $Cr(r)$, to be identified experimentally as the center of a dark triangle in Fig. 1A and all equivalent T($r$). The measured spatial density of such sites in our samples is 0.23 nm², indicating ~8% Cr/Bi substitutions (*SI* section 2). Figure 1C then shows an image of the differential conductance $g(r, E = -50meV)$ measured in the same field of view (FOV) as Fig. 1A, with the inset showing its power-spectral-density Fourier transform g($q$,E) which exemplifies the surface state quasiparticle scattering interference (QPI) phenomena (*31,32,33*).

Next we introduce the QPI technique to the study of ferromagnetically gapped TI surface states, by simultaneously imaging the tunnel current $I(r, E = eV)$ and $g(r, E)$. We use this approach because the density of surface electronic states $N(r, E)$ is related to the differential tunneling conductance as $g(r, E) \propto \left[eI_s / \int_0^{eV_s} N(r, E')dE'\right] N(r, E)$ ($I_s$ and $V_s$ are arbitrary parameters) . Thus valid determination of $N(r, E)$ is not possible because the denominator $\int_0^{eV_s} N(r, E')dE'$ is unknown and heterogeneous (Ref. 24 and see below). We mitigate the consequent and serious systematic errors by using the function $K(r, E) = \langle I(E) \rangle g(r, E)/I(r, E)$ because $I^{-1}(r, E) \propto \int_0^{eV_s} N(r, E')dE'$ and $\langle I(E) \rangle$ provides normalization from the spatially averaged current. Then, when $g(r, E)$ and $I(r, E)$ are measured at T=4.5K in the FOV of Fig. 1A and $K(q, E)$, the power-spectral-density Fourier transform of $K(r, E)$, is determined, the results are shown in supplementary movie #1. Analyzing these QPI data for E<E$_F$, we find dispersion of the surface states consistent with ARPES studies of the same samples (*SI* section 3). Moreover, above $E_F$ the QPI data reveal vividly the appearance of the Dirac-mass gap starting at $E \approx 130$ meV (*SI* section 3). Figures 2A-P illustrate this result directly using a sequence of typical $K(\boldsymbol{q}, E)$ images that span the energy range 105meV<E<220 meV. With increasing E, the surface-state QPI signature evolves smoothly and with diminishing |q| until ***q***=0 is reached just above $E \approx 130 meV$ (Fig. 2G). At this point, the surface-state QPI disappear leaving only electronic noise near ***q***=0. Just below $E \approx 200 meV$, the surface-state QPI signatures reappear once again, emerging from ***q***=0 (Fig. 2M). In Fig. 2Q we show the measured $K(q_y, E)|_{q_x=0}$ from the same $K(\boldsymbol{q}, E)$ data revealing directly how the QPI dispersion evolves towards ***q***=0 for $E < 130 meV$, disappears $130 meV < E < 200 meV$, and then reappears to evolve away from ***q***=0 at $E > 200 meV$. This situation is very well described by two surface state bands $E_{\pm}(k) = E_D \pm \sqrt{(\hbar v)^2 k^2 + \Delta^2}$ meaning that the energy range devoid of QPI between the two band edges is twice the Dirac-mass gap Δ. For comparison, the spatially averaged differential conductance $\bar{g}(E)$ in the same FOV is shown in Fig. 2R. Its magnitude becomes indistinguishable from zero between $130 meV < E < 200 meV$, demonstrating independently that the Dirac-mass gap Δ has opened in this range as indicated by the arrows spanning 2Δ. Thus, the magnitude of Δ can be detected both directly and locally by measuring half the energy range where tunneling conductance is indistinguishable from zero in g(***r***,E) (e.g. black arrows Fig. 2R). These, and equivalent observations in multiple samples, also demonstrate that the ungapped Dirac point is somewhere near $E_D$=+150meV. More significantly they also show directly that the Dirac-mass gap magnitude is Δ~30meV, and that the bulk states also seem gapped because no tunneling is detected at T=4.5K in this energy range. Thus, as widely reported (*25-28*), the $Cr_x(Bi_{0.1}Sb_{0.9})_{2-x}Te_3$ materials appear to be excellent candidates to exhibit the exotic new phenomena predicted for the gapped surface states of a TI.

Next we introduce the Dirac-mass 'gapmap' technique designed to measure spatial arrangements of $\Delta(\boldsymbol{r})$, and apply it throughout the FOV of Fig. 1A. Atomically-resolved $g(\boldsymbol{r}, E)$ data are measured at 4.5K and, for each pixel location ***r***, we define a mask function $f(\boldsymbol{r}, E) = 1$ if $g(\boldsymbol{r}, E) < 40$ pS (the tunnel conductance noise floor) and $f(\boldsymbol{r}, E) = 0$ otherwise. This determines the value of Dirac-mass gap $\Delta(\boldsymbol{r}) = \frac{1}{2}\int f(\boldsymbol{r},E)\,dE$. Figure 3A shows a sequence of spectra, $\bar{g}(\Delta)$, each representing the average of all spectra measured to have the same value of $\Delta$. Each $\bar{g}(\Delta)$ is shifted upwards from the others by the same amount for clarity; the zero of conductance in each case is indicated by a fine horizontal line. The value of $2\Delta$ is then indicated for each $\bar{g}(\Delta)$ by the energy span between the pairs of arrows in Fig. 3A. Figure 3B shows the histogram of all values of $\Delta(\boldsymbol{r})$ detected in the FOV of Fig. 1A and labels each value of $\Delta$ using a color scale. The distribution of $\Delta(\boldsymbol{r})$ is centered near $\Delta$=28 meV, exhibits a wide but approximately normal distribution. Finally, Figure 3C shows the atomically resolved spatial arrangements of $\Delta(\boldsymbol{r})$ in a Dirac-mass 'gapmap'. The autocorrelation width of this image is 1.24 nm (*SI* section 4), indicating that there are a wide variety of nano-domains of like $\Delta$, each with radius near 0.62 nm. Due to coalescence, of course, many regions of similar $\Delta$ are significantly larger (Fig. 3C).

Importantly, because $E_F$ is ~150 meV below the implied Dirac point, the hexagonal warping of the Fermi surface (e.g. Fig. 1C, *SI* section 1) should play a significant role, because the magnitude of $k_F$ becomes a function of direction in momentum space. A rotational anisotropy in the Dirac-mass gap would then be expected from models similar to Eqn. 1, if dopant atoms interact via the TI surface states. Taking the Fourier transform of the measured $\Delta(\boldsymbol{r})$ (Fig. 3C) this is what we detect (Fig. 3D), with the lobes in $\Delta(\boldsymbol{q})$ oriented in the expected directions. When viewed in much larger field of view (Fig. 3E), the $\Delta(\boldsymbol{r})$ fluctuations remain rather uniformly distributed with no extreme outliers in the mass gap fluctuations. Moreover, there is an energy shift of each local Dirac point $E_D(\boldsymbol{r})$ as deduced from the g($\boldsymbol{r}$,E) spectra (*SI text*, section 5). This varies weakly in space (Fig. 3F) with ~10 meV fluctuations of $E_D$ (inset Fig. 3F). These relatively minor acceptor-induced band shifts will not alter the Fermi surface appreciably and therefore negligibly impact the primary electronic processes leading to the Dirac-mass gap, whose characteristic energy range is at least 5 times larger. Overall, these data reveal for the first time how strikingly disordered at the nanoscale (Fig. 3) are the Dirac-mass gaps of doped ferromagnetic TIs.

Could the type of Dirac-mass disorder shown in Figs 3C,E be driven by local variations of $M_z(r)$ due to deviations in $n(r)$ arising from the random distribution of the Cr dopant atoms? To study this issue, we first identify the location of each Cr atom $Cr(r) = \delta(r - r_{Cr})$ where $r_{Cr}$ are the centers of all dark triangles in Fig 1A. The results are indicated by the red triangles in Fig. 4. Next, in order to establish a local measure of areal density, $n(r)$, it is necessary to define a distance scale (*SI text*, section 4). To do so we define two images $n(r,\xi)$ and $\Delta(r,\zeta)$ where $\xi(\zeta)$ is the Gaussian correlation length of $n(r)(\Delta(r))$ images. Then we identify the maximum in the normalized cross correlation of these two processed images $n(r,\xi) : \Delta(r,\zeta)$ as a function of both $\xi$ and $\zeta$. We find that it occurs at $\xi \sim 0.82 \pm 0.1$ nm, the empirical radius of influence of each Cr atom at which their arrangements correspond maximally to the Dirac-mass arrangements in $\Delta(r)$, while for $\Delta(r)$ this occurs at $\zeta \sim 0.55 \pm 0.1$ nm. Notwithstanding which microscopic interactions drive the surface ferromagnetism in CBST, we find that the distribution of Cr atoms $n(r)$ is correlated manifestly with the $\Delta(r)$. One can see this directly in Fig. 4, which is a representative subset of Fig. 3C on which every Cr site is represented by a red triangle. Furthermore, the inset shows the plot of average value of $\Delta(r)$ in Fig. 4 associated with each value of $n(r)$, and reveals a quasi-linear relationship between local Dirac-mass gap and local Cr density. Indeed, the slope of $\Delta(r) = \lambda \cdot n(r)$ allows the surface-state-dopant interaction energy scale (e.g. Eqn. 2) to be measured directly. For Cr-doped CBST, we find that $J^* = 145 \pm 25$ meV·nm$^2$ everywhere, despite the strong fluctuations in $n(r)$ (Figs 1A,4).

To summarize: by studying recently developed materials $Cr_{0.08}(Bi_{0.1}Sb_{0.9})_{1.92}Te_3$ we provide simultaneous visualization of the location of magnetic dopant atoms and the Dirac-mass gap in a ferromagnetic TI. Imaging $\Delta(r)$ reveals that its nanoscale disorder is intense as anticipated (*1-6*). Next, we directly demonstrate that the $\Delta(r)$ disorder is caused by fluctuations in the areal density $n(r)$ of magnetic Cr atoms in the crystal termination layer. And despite the nanoscale heterogeneity of Dirac-mass, we confirm the anticipated relationship $\Delta(r) \propto n(r)$ throughout, indicating a universal dopant/surface-state interaction energy scale $J^* = 145$ meV·nm$^2$. The Dirac-mass gapmap technique introduced here also reveals what appear to be $\Delta$ domains (e.g. Figures 3,4) and therefore may be extended, in future, to visualize the chiral states expected at the perimeters of FM domains (*34*). Finally, these data demonstrate that, to achieve the exotic physics expected of time-reversal-symmetry breaking TI materials (*7-18*), an approach to controlling the severe dopant-induced Dirac-mass gap disorder will first need to be identified. Application of techniques initiated here provides a promising new approach to this important challenge.

**Acknowledgements**

We acknowledge and thank A. V. Balatsky, J. Checkelsky, M. Franz, Z. Hasan, A. P. Mackenzie, V. Madhavan, and J. W. Orenstein, and very helpful discussions and communications. Experimental studies and sample fabrication were supported by US DOE under contract number DE-AC02-98CH10886 (I.L., J.C.D., T.V., J.T & G.G.), and under FlucTeam program at Brookhaven National Laboratory Contract DE-AC02-98CH10886 (S.B. and C.K.K). J.L. acknowledges support from the Institute of Basic Science of Korea under IBS-R009-D1. J.C.D. also acknowledges the support for conceptual design studies for this project with J.P. Reid under EPSRC Programme Grant 'Topological Protection and Non-Equilibrium States in Correlated Electron Systems".

**Figure Captions**

**Fig. 1. Cr dopant-atom locations at the TI surface**

(**A**) Topographic image T(*r*) of $Cr_{0.08}(Bi_{0.1}Sb_{0.9})_{1.92}Te_3$ surface in a 47x47 $nm^2$ FOV. **Inset:** Zoomed-in topographic image of red dashed box area. Both images were measured at 10pA/-200meV. A single Cr dopant atom exists substituted at the Bi/Sb site at the symmetry point of every dark triangle in T(*r*) as discussed in (B).

(**B**) Schematic of the crystal structure of $Cr_x(Bi_{0.1}Sb_{0.9})_{2-x}Te_3$. Each substitutional Cr atom is located at a Bi/Sb site at the symmetry point of a triangle of surface Te atoms (A).

(**C**) Measured differential conductance g(*r*,E=-50 meV) in same field of view (FOV) as (A). **Inset:** Power-spectral-density Fourier transform of the *r*-space differential conductance image showing the q-space signature of Friedel oscillations due to scattering interference of the surface state electrons.

**Fig. 2. Measuring Dirac-mass Gap from both Tunneling Spectrum and QPI**

(**A-P**) Quasiparticle interference of TI surface states is visualized as a function of energy using $K(q,E)$, the Fourier transform of $K(r,E) = \langle I(E) \rangle g(r,E)/I(r,E)$. The complete data $K(q,E)$ are provided in movie format. Here we see directly the disappearance of the surface state QPI in an energy window between 130meV and 200 meV.

(**Q**) Measured dispersions in quasiparticle interference of the TI surface states is plotted using $K(q_x = 0, q_y, E)$, the *E-q* line cut along $\bar{\Gamma} - \bar{K}$. Here again the evolution of scattering interference signature of surface states to reach *q*=0 at ~130meV, followed by their disappearance, and the reappearance at and dispersion away from *q*=0 near E~200meV, is manifest. The Dirac gap magnitude Δ is half the energy range between the two *q*=0 tips of the surface state bands, as indicated.

(R) $\bar{g}(E)$, the spatially averaged tunneling conductance, simultaneously measured with (Q) showing that conductance becomes indistinguishable from zero within the same energy window as in (Q). Again, this indicates that the Dirac gap magnitude Δ is half the energy range between points at which conductance disappears/reappears, as indicated.

**Fig. 3. Dirac-Mass Gapmap**

(A) Measured conductance spectra, $\bar{g}(\Delta)$, each representing the average of all spectra with the same value of Δ from the FOV of Fig. 1A. Each $\bar{g}(\Delta)$ is shifted upwards for clarity and the zero of conductance is shown by a fine horizontal line. 2Δ in each $\bar{g}(\Delta)$ is the energy span between the pairs of arrows.

(B) Histogram of the $\Delta(\boldsymbol{r})$ measured in the FOV of Fig. 1A.

(C) Dirac-mass gapmap $\Delta(\boldsymbol{r})$ (or Dirac-mass map $\mathrm{m}(\boldsymbol{r})$) extracted from $g(\boldsymbol{r}, E)$ measured in the FOV of Fig. 1A. This is typical of maps made using similar parameters on multiple samples of this compound. Tip-induced band bending effects have been systematically ruled out by checking that these results are independent of the tip elevation.

(D) Fourier transform of the Dirac-mass gap map $\Delta(\boldsymbol{r})$ from the FOV of Fig. 1A. The *q*-space anisotropy in $\Delta(\boldsymbol{q})$ is as would be expected due to the anisotropic values of $k_F$ in the TI of the surface states.

(E) Dirac-mass gap $\Delta(\boldsymbol{r})$ measured in the 360x360 nm$^2$ FOV much larger than the map of (C). **Inset**: The histogram of $\Delta(\boldsymbol{r})$.

(F) Map of estimated ungapped Dirac-point energy (gap center), $E_D(\boldsymbol{r})$ $(= \int f(\boldsymbol{r}, E) \cdot E \, dE / \int f(\boldsymbol{r}, E) \, dE)$ obtained in the same FOV as (E), where $f(\boldsymbol{r}, E)$ is the gap mask function defined in the text. **Inset**: Each data point represents the average value of gap center $E_D$ over all the regions having the same value of Cr density *n*.

**Fig. 4. Atomic-scale Measurements of Interaction Strength of Surface-states with Magnetic Dopant Atoms**

Measured Dirac-mass gap map $\Delta(\boldsymbol{r}, \zeta = 0.55\ nm)$ with Gaussian smoothing length $\zeta$, overlaid with Cr locations measured from Fig. 1(A) (red triangles). Cr atoms are observed to be positioned with high probability in the larger gap areas (yellow), but rarely in the smaller gap areas (blue). The other fainter features in topography (white) are shown not to occur at a Bi/Sb substitutional sites, so we do not assign them as magnetic dopant atoms. **Inset**: Each data point represents the average value of Dirac mass gap over all the regions having the same value of Cr density *n*. The resulting slope of best linear fit yields *J**=145 meV·nm². This is the first atomic-scale measurement of the interaction strength of surface-states with magnetic dopant atoms in a ferromagnetic TI. The uncertainty represented by two dashed lines here is not statistical but comes from the systematic uncertainty in magnitude of $S_z$, which we take to be 20%.

Figure 1

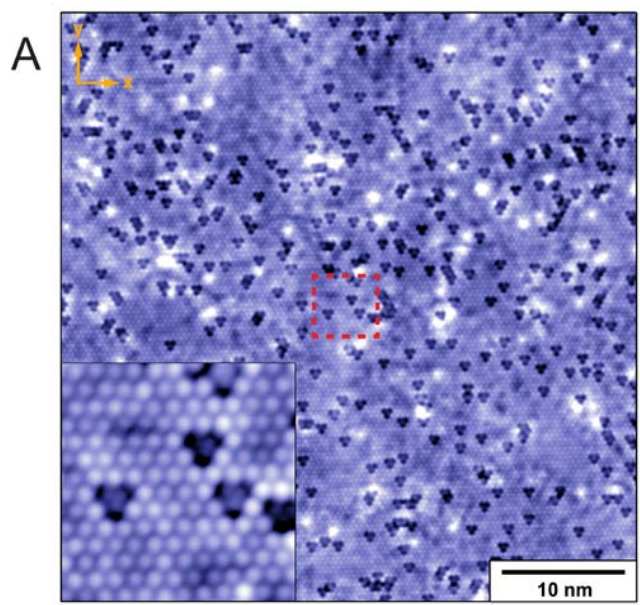

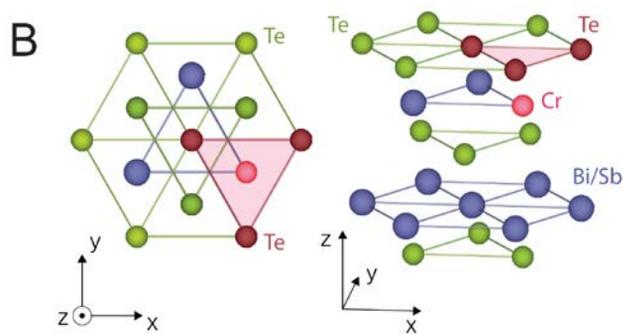

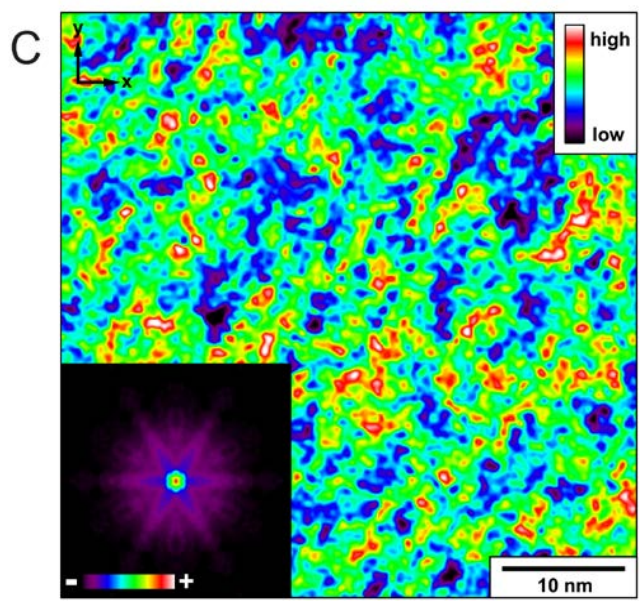

Figure 2

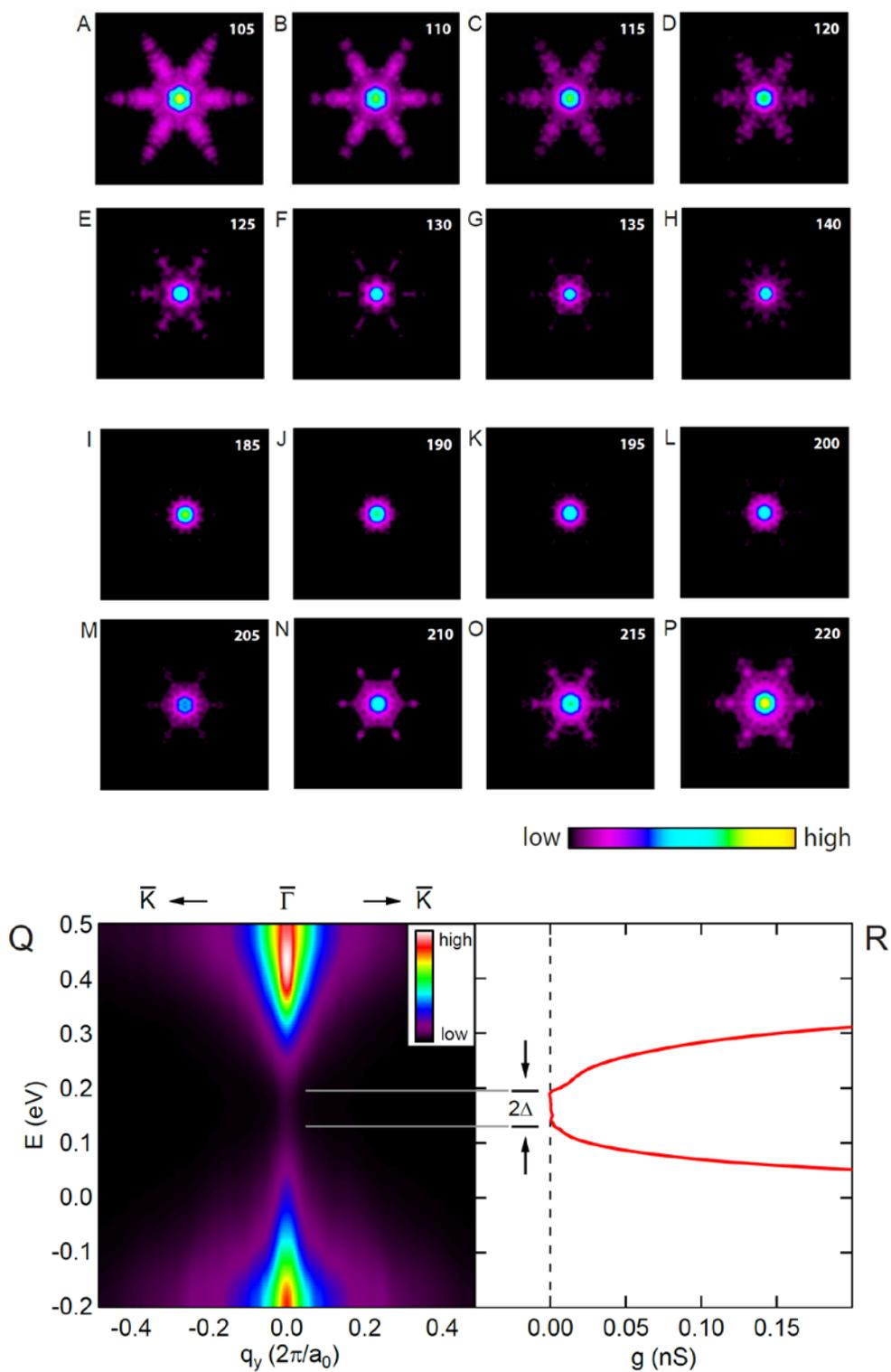

Figure 3

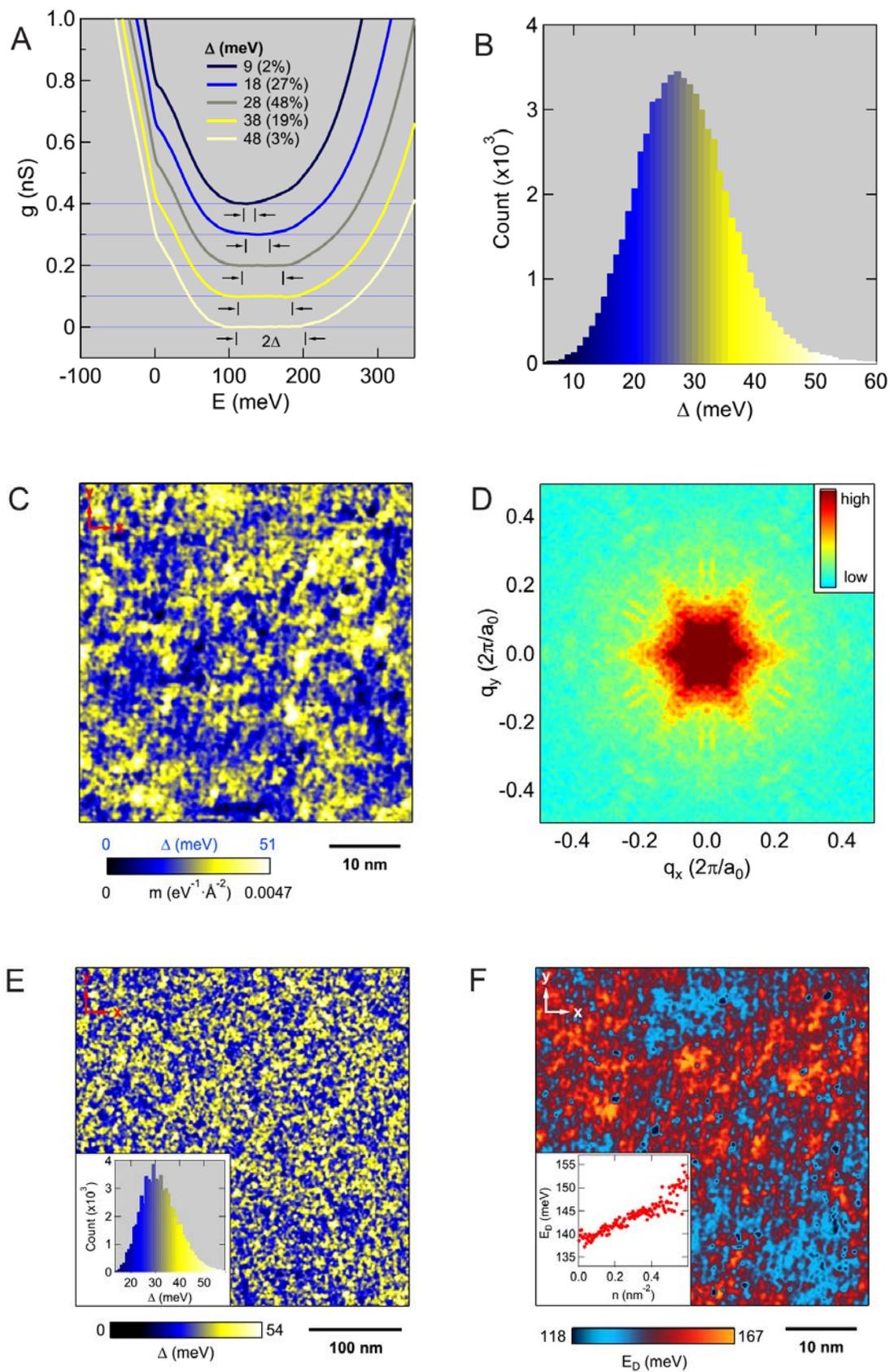

Figure 4

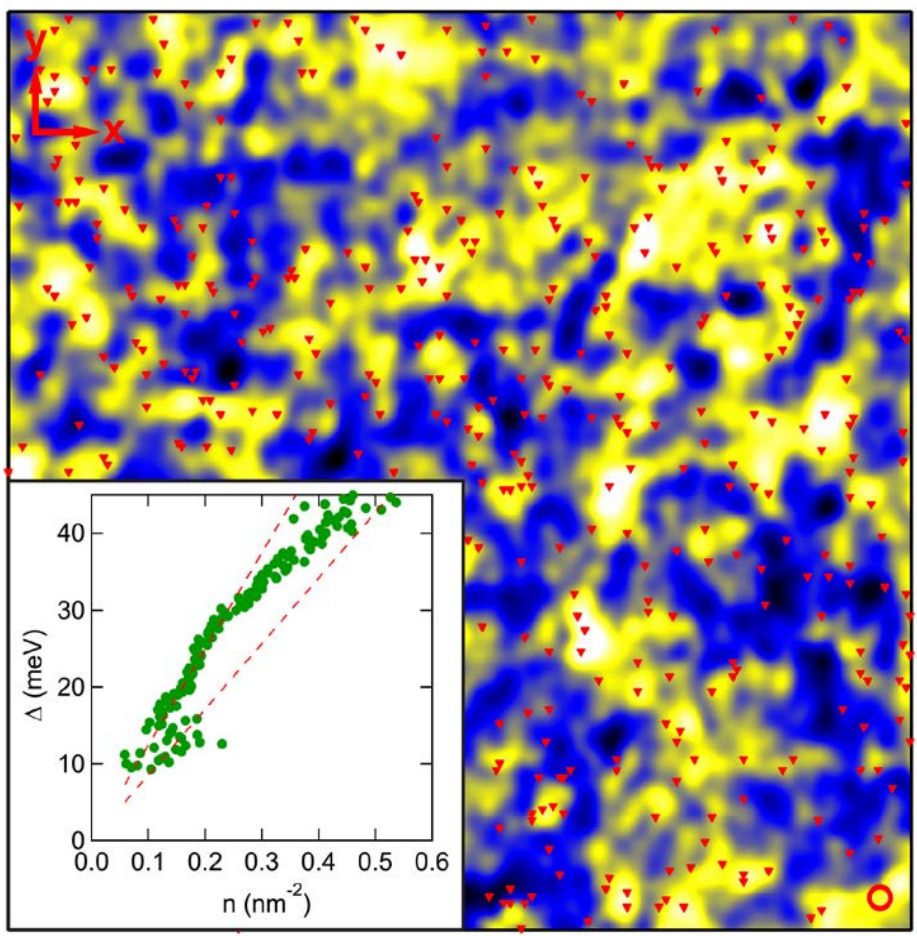

# Supporting Information

# Imaging Dirac-Mass Disorder from Magnetic Dopant-Atoms in the Ferromagnetic Topological Insulator $Cr_x(Bi_{0.1}Sb_{0.9})_{2-x}Te_3$

Inhee Lee, Chung Koo Kim, Jinho Lee, S. J. L. Billinge, R. D. Zhong, J. A. Schneeloch, T. S. Liu, T. Valla, J. M. Tranquada, G. D. Gu, and J. C. Davis

# 1. Dirac-mass Disorder in Ferromagnetic Topological Insulators

An important question about ferromagnetic TI's is how (or whether) the spatial arrangements of the magnetic dopant atoms influence the homogeneity of Dirac mass gap. Elementary theoretical models for this situation are based upon the hypothesis that pairs of magnetic-dopant spins on the TI sample surface can have a component of interaction that is Dirac surface-state mediated. The coupling strength for two magnetic spins $\Phi_Z(r)$ is then given, for example, by Eq. 1B in the main text. As a concrete example we consider an hypothetical arrangement of magnetic dopant atoms in the termination layer of a ferromagnetic TI with an areal dopant density $n_0 \approx 0.25$ nm$^{-2}$ with average nearest-neighbor distance $a_m \approx 1$ nm (Fig. S1A). Only the z-component (normal to the sample surface) of magnetism is considered. Figure S1B then shows $\Phi_Z(r)$ for a typical Dirac surface-state band structure with $k_F \approx 0.06\ (\pi/a_0)$; $\Phi_Z(r)$ oscillates with a wavelength $\lambda = \pi/k_F$ alternating its sign as a function of $r$ so that two dopant spins tend to align in the same (opposite) direction for $\Phi_Z(r) > 0$ ($\Phi_Z(r) < 0$). As shown in the inset $\Phi_Z(a_m) > 0$ indicates that magnetic-dopant spins would favor ferromagnetism for $a_m$ and $k_F$ with the above values. Estimating the self-consistent magnetic properties induced by the interplay of surface-states with randomly distributed magnetic dopant atoms can be achieved by starting with an initial polarization $\bar{S}_z$ for all magnetic spins and then calculating the local magnetic field $B_z$ at each dopant position $\boldsymbol{r}_i$ generated by all other dopant spins at $\boldsymbol{r}_j$ ($\neq \boldsymbol{r}_i$) through a RKKY interaction $\Phi_Z(|\boldsymbol{r}_i - \boldsymbol{r}_j|)$ in a model such as described by Eq. 1B. Given this $B_z(\boldsymbol{r})$ one can then calculate the spin polarization $S_z(\boldsymbol{r})$ with the following relation (1)

$$S_z(\boldsymbol{r}) = S \cdot B_S\left(\frac{B_z(\boldsymbol{r}) \cdot S}{k_B T}\right) \tag{S1}$$

where the Brillouin function $B_S(x)$ is

$$B_S(x) = \frac{2S+1}{2S}\coth\left(\frac{2S+1}{2S}x\right) - \frac{1}{2S}\coth\left(\frac{x}{2S}\right) \tag{S2}$$

and assuming $S = 3/2$ as spin number. Then, the $\bar{S}_z$ for individual dopant-atom spins calculated from $S_z(\boldsymbol{r})$ is inserted again into Eq. 1A in order to get a new $B_z(\boldsymbol{r})$; this feedback process is repeated until an unchanging and self-consistent $\bar{S}_z$ is achieved. For above magnetic dopant distribution (Fig. 1A) and parameters including a self-consistent $\bar{S}_z=1.45$, the final results of this procedure for $S_z(\boldsymbol{r})$ and $B_z(\boldsymbol{r})$ are shown in Fig. S1C and S1D, respectively. From $S_z(\boldsymbol{r})$ in Fig. S1C, we then model the heterogeneous magnetization $M_z(\boldsymbol{r})$ by convolving with a Gaussian of correlation length $\xi_M$ of the ferromagnetism

$$M_z(\boldsymbol{r}, \xi_M) = C \cdot S_z(\boldsymbol{r}) * G(\boldsymbol{r}, \xi_M) \quad \text{(S3A)}$$

$$G(\boldsymbol{r}, \xi_M) = \frac{1}{\pi \xi_M^2} e^{-r^2/\xi_M^2} \quad \text{(S3B)}$$

Here $C$ is the constant yielding an average magnetization $\bar{M}_z = n_0 \bar{S}_z \mu_B = 0.33$ $\mu_B$/nm$^2$, $S_z(\boldsymbol{r})$ is the map of spin polarization (e.g. Fig. S1C) and the notation * represents convolution. Figure S1E shows such a result for $M_z(\boldsymbol{r})$ if $\xi_M \approx 0.80$ nm. Lastly, one could estimate the spatial structure of the Dirac-mass gap $\Delta(\boldsymbol{r})$ induced by an heterogeneous $M_z(\boldsymbol{r})$ by applying Eq. 2 of the main text. This relation can be made more realistic by considering the effective correlation length of coupling between Dirac surface states and $M_z(\boldsymbol{r})$ such as

$$\Delta(\boldsymbol{r}, \zeta_D) = \frac{J^* |M_z(\boldsymbol{r})|}{2\mu_B} * G(\boldsymbol{r}, \zeta_D) \quad \text{(S4)}$$

where $G(\boldsymbol{r}, \zeta_D)$ and * are defined as Eq. S3. Figure S1F then shows the result for $\Delta(\boldsymbol{r})$ if $\zeta_D = 1$ nm. Thus, elementary models for magnetically doped ferromagnetic TI materials indicate that the surface magnetism $M_z(\boldsymbol{r})$ and the resulting $\Delta(\boldsymbol{r})$ could be heterogeneous on the nanoscale if the dopant atoms are randomly distributed with an areal density n($\boldsymbol{r}$). More generally, as any heterogeneous surface $M_z(\boldsymbol{r})$ should influence $\Delta(\boldsymbol{r})$ in this fashion, it is important to determine empirically the structure of $\Delta(\boldsymbol{r})$ in real materials.

## 2. Determination of the chromium concentration

Precise knowledge in a given crystal of the Cr doping level is key information for the proper characterization of the magnetic properties of the $Cr_x(Bi_{0.1}Sb_{0.9})_{2-x}Te_3$ (CBST) samples. SI-STM can visualize the non-uniform spatial distribution of Cr atoms in real space for direct determination of their locations, as well as for density determination by simple counting. Typical topographic images of our cleaved CBST surfaces, as shown in Fig. 1A of the main text and in Fig. S2, exhibit characteristic dark triangles due to the presence of the Cr dopant atom at a Bi/Sb site immediately underneath the top Te layer (*2*). One can manually count the number of these triangles to obtain the value $x$ in the chemical formula under two assumptions – 1) only the dark triangles are due to the Cr atoms in the Bi/Sb layer closest to the surface; Cr atoms in the Bi/Sb layer *second next* to the surface are unobservable, 2) two Bi/Sb layers within a quintuplet have equal Cr concentration. Fig. S2 presents one example of actual counting where 26 triangles are individually marked with yellow arrows. Here since 10nm×10nm field-of-view (FOV) contains approximately 635 atoms per layer, the Cr density is estimated to be 4.09% per layer, or equivalently, $x{\sim}8.18\%$. The same calculation gives the statistically equivalent $x{\sim}7.52\%$ when applied to the FOV in Fig. 1A of the main text.. The effective chemical formula appropriate for the samples studied in this work is thus $Cr_{0.08}(Bi_{0.1}Sb_{0.9})_{1.92}Te_3$.

## 3. Quasiparticle Interference (QPI) Imaging and Determination of Δ

In this section we determine the average Dirac mass gap Δ by means of QPI imaging analysis. In general, QPI data measured when using constant current topography to determine the tip-surface distance for each junction have a serious systematic error often referred to as the 'setup effect' (*3*). This occurs universally when there is a heterogeneous electronic structure; it unavoidably leads to the distortion of actual QPI images. Therefore, in order to remove this effect, we apply a slightly modified Feenstra method on our QPI data (*4*) by utilizing a new image

$$K(\boldsymbol{r}, E) = \frac{g(\boldsymbol{r}, E)}{I(\boldsymbol{r}, E)} \cdot \bar{I}(E) \tag{S5}$$

where $g(r, E)$ is the unprocessed differential conductance map, $I(r, E)$ is the simultaneously measured current map, and $\bar{I}(E)$ is the averaged value of $I(r, E)$. The constant $\bar{I}(E)$ is just for preserving the relative intensities of $g(r, E)$ between $E$ layers by canceling out $I(r, E)$ in the division.

Movie S1 shows our typical $Cr_{0.08}(Bi_{0.1}Sb_{0.9})_{1.92}Te_3$ g(**r**,E) and its corresponding K(**q**,E) processed by Eq. S5 (left and right frames, respectively). As energy $E$ increases from -200 meV, the average local density of states $\bar{g}(r, E)$ diminishes and at the same time the overall QPI signatures also get weaker and shrink toward the center of **q**-space exhibiting the expected dispersion of topological Dirac surface states. By E~130 meV, both g(**r**,E) and K(**q**,E) are totally suppressed implying the surface states are entering the Dirac gap regime (a weak **q**~0 core remains at the center of K(**q**,E) due to long wavelength noise). This QPI suppression to zero continues up to E~200 meV, the upper edge of the Dirac gap (see Fig. 2 for the high contrast K(**q**,E) stills). Above 200 meV, the g(**r**,E) and K(**q**,E) intensify and disperse once again as they exit the Dirac gap. From this QPI analysis and the simultaneously measured average spectrum (Fig. 2), we find that the Dirac mass gap exists in 130 - 200 meV for these $Cr_{0.08}(Bi_{0.1}Sb_{0.9})_{1.92}Te_3$ samples.

For simplicity, we determine the isotropic (minimum) $q_F$ from the QPI data, thus neglecting the much longer $q_F$ that are of less relevance to any surface-state mediation of interactions. Thus, we first take the azimuthally averaged $K(q, E = 0 \text{ meV})$ shown in Fig. S3A, and fit it with

$$K_f(q) = A(q)(= A_0 e^{-q^2/w_A^2}) + B(q)(= B_0 e^{-q^2/w_B^2}) + C_0 \tag{S6}$$

consisting of two Gaussian functions as shown in Fig. S3C. By eliminating $B(q)$, the QPI $q_F = 0.066\ (2\pi/a_0)$ can be attained from the Gaussian width $2w_A$ of $A(q)$ as shown in Fig. S3C. The Fermi wavenumber $k_F = q_F/2 = 0.067(\pi/a_0)$ is then determined and the Fermi velocity $v$ derived from the following dispersion relation

$$E_\pm(\mathbf{k}) = E_D \pm \sqrt{(\hbar v)^2 k^2 + \Delta^2}. \tag{S7}$$

We obtain $v = 3.3 \pm 0.01$ eV·Å for $E_D = 165$ meV and $\Delta = 35$ meV which are determined from the QPI results described above. The **q**-dispersion estimated by the simple band model in Eq. S7 with these parameters is shown as white curves in Fig. S4B. They are in good agreement with background images, $E$-$q$ line cuts of $K(\mathbf{q}, E)$ along $\bar{\Gamma} - \bar{M}$ and $\bar{\Gamma} - \bar{K}$, respectively. Furthermore, this $v$ can be used in the conversion of gap $\Delta$ into mass $m$ with the relation.

$$m = \frac{\hbar^2}{\left(\frac{\partial^2 E}{\partial k^2}\right)}\bigg|_{k=0} = \frac{\Delta}{v^2} \tag{S8}$$

We have also carried out Angle Resolved Photoemission Spectroscopy (ARPES) measurements with $Cr_{0.08}(Bi_{0.1}Sb_{0.9})_{1.92}Te_3$ samples from the same original crystal. Fig. S4C shows the ARPES spectrum image in which the surface state band is dispersing from -0.2 eV to 0 eV ($E_F$) along the trajectory of the maxima of lorentzian fit indicated by yellow dots. For comparison, the QPI dispersion ($q = 2k$) expected from the ARPES measurement is overlaid with our QPI data in Fig. S4B. Dispersions of surface state from our QPI (white curve) and ARPES (yellow dots) show good agreement (within the mutual uncertainties due to using different techniques at different temperatures) in Fermi velocity $v_F$, 3.24 and 3.3eV·Å, respectively, which are estimated from the slope of $E$ and $q$ below $E_F$. These data support ARPES as a valid technique for surface-state band determinations of ferromagnetic TI when $E<E_F$.

## 4. Spatial Correlations

We use spatial correlations to estimate characteristic length scales in the electronic structure images (Figs 3,4). First, the correlations within the Dirac-mass gap $\Delta(\mathbf{r})$ can be examined by its normalized auto-correlation $AC_\Delta(\mathbf{r})$, where

$$AC_f(\mathbf{r}) = \frac{\int [f(\mathbf{r}') - \bar{f}] \cdot [f(\mathbf{r}' + \mathbf{r}) - \bar{f}] \, d\mathbf{r}'}{\int [f(\mathbf{r}') - \bar{f}]^2 \, d\mathbf{r}'} \tag{S9}$$

and $\bar{f}$ is the average value of the image $f(\mathbf{r})$. Fig. S5A shows $AC_\Delta(\mathbf{r})$ for $\Delta(\mathbf{r})$ in Fig. 3C, and Fig. S5B shows its azimuthally averaged curve. The correlation length has a width $w = 0.62 \pm 0.011$ nm which is obtained from the Lorentzian fit $L(r) = L_0/(r^2 + w^2)$ of an azimuthally averaged curve.

The interplay of the observed magnetic dopant density $n(\mathbf{r})$ and the observed Dirac mass gap $\Delta(\mathbf{r})$ are explored in this way. We define $n(\mathbf{r}, \xi)$ and $\Delta(\mathbf{r}, \zeta)$ by assigning two correlation lengths $\xi$ and $\zeta$ to $Cr(\mathbf{r}) = \delta(\mathbf{r} - \mathbf{r}_{Cr})$ (e.g dark triangles in Fig. 1A or red triangles in Fig. 4) and $\Delta(\mathbf{r})$ (Fig. 3C), where convolutions with normalized Gaussian functions, $\frac{1}{\pi\xi^2}e^{-r^2/\xi^2}$ and $\frac{1}{\pi\zeta^2}e^{-r^2/\zeta^2}$ are implemented, respectively. Then, two independent parameters $\xi$ and $\zeta$ are adjusted to find the maximum of $XC_{n:\Delta}(\xi, \zeta)$ where the normalized cross-correlation between two images, $f(\mathbf{r})$ and $g(\mathbf{r})$, is defined by

$$XC_{f:g}(\mathbf{r}) = \frac{\int [f(\mathbf{r}') - \bar{f}] \cdot [g(\mathbf{r}' + \mathbf{r}) - \bar{g}] \, d\mathbf{r}'}{\sqrt{\int [f(\mathbf{r}') - \bar{f}]^2 \, d\mathbf{r}' \cdot \int [g(\mathbf{r}') - \bar{g}]^2 \, d\mathbf{r}'}} \quad (S10)$$

At maximum of $XC_{n:\Delta}(\xi, \zeta)$ (white cross in Fig. S5C), we find $\xi_n = 0.82 \pm 0.09$ nm and $\zeta_\Delta = 0.55 \pm 0.09$ nm as correlation lengths for fluctuation of Cr dopant density and mass gap domain, respectively.

## 5. Dirac Point Disorder

Here we estimate the spatially non-uniform distributions of the Dirac point $E_D(\mathbf{r})$ detected in our $K(\mathbf{r}, E)$ data. In the simple band model given by Eq. S7, the Dirac point is positioned at the center of the Dirac-mass gap in the spectrum. The measured local values of the gap center $E_D(\mathbf{r})$ are reported in Fig. 3F. Fig. S6A shows the histogram of all values of $E_D(\mathbf{r})$ in Fig. 3F with each value of $E_D$ represented by the same color scale used in Fig. 3F.

The normalized auto-correlation of $E_D(\mathbf{r})$ is shown in Fig. S6B. In order to find the correlation length $\xi$ of fluctuating Cr dopant density associated with Dirac point disorder, we perform the cross correlation of $E_D(\mathbf{r})$ and $n(\mathbf{r}, \xi)\left(= Cr(\mathbf{r}) * \frac{1}{\pi\xi^2}e^{-r^2/\xi^2}\right)$ in a similar way described in SI Section 4. Now, only $\xi$ is adjusted to find the maximum of $XC_{n:E_D}(\xi)$ because of no local correlation in $E_D(\mathbf{r})$ (Fig. S6B) where $XC_{n:E_D}(\xi)$ is defined in Eq. S10. As shown in Fig. S6C, we find $\xi_n = 1.37 \pm 0.09$ nm.

Then to examine the effect of Cr dopants on the Dirac point $E_D$, we sort the values of $E_D(\mathbf{r})$ together with $n(\mathbf{r}, \xi_n)$ at the same location, and then plot the average of this sorted $E_D$ as a function of the average of its associated $n$. Its result is shown in the inset of Fig. 3F which exhibits the straightforward positive correlation between Cr dopant density and Dirac point $E_D$, i.e. Cr dopants induce the shifting of Dirac point $E_D$ to the higher energy as expected for Cr acceptor atoms, but only by about 10 meV at most. This weak band shifting should not affect the Fermi wavevector substantially, nor the Dirac mass gap itself since its characteristic energy range of the gap is at least 5 times larger.

## 6. Materials and Methods

Single crystals with nominal composition $Cr_{0.15}(Bi_{0.1}Sb_{0.9})_{1.85}Te_3$ were grown by a modified floating-zone method. The elements of high purity (99.9999%) Bi, Sb, Cr, and Te were loaded into double-walled quartz ampoules and sealed under vacuum. The materials first were melted at 900 °C in a box furnace and fully rocked to achieve homogeneous mixture. The 12 mm diameter pre-melt ingot rod in a quartz tube were mounted in a floating-zone furnace. In the floating-zone furnace, the pre-melt ingot rods were first pre-melt at a velocity of 200 mm/hr and then grown at 1.0 mm/hr in 1 bar Ar atmosphere. Because the segregation coefficient of indium is less than 1, the Cr contained in the feed material would then prefer to remain in the liquid zone. As a result, a homogeneous Cr concentration along the whole grown rod is difficult to achieve. The Cr concentration in the as-grown single crystals is thus less than the normal concentration in the feed rod.

Overall magnetic properties of the samples used in our SI-STM experiments are evaluated using SQUID magnetometry as shown in Fig. S7. Temperature (Fig. S7A) and field (Fig. S7B) dependence of the magnetization exhibit the common ferromagnetic features of the samples with bulk Curie temperature $T_c$ ~ 18 K and coercive field $H_c$ ~ 15 mT at $T$=4.5 K. These SQUID measurements are carried out in the direction $z$ perpendicular to the plane of the sample.

Our $Cr_x(Bi_{0.1}Sb_{0.9})_{2-x}Te_3$ samples are cleaved in ultra-high-vacuum environment below T=10K, and then immediately inserted into the STM head for spectroscopic measurements at T=4.5K. The standard lock-in technique was used to obtain differential tunneling conductance data $dI/dV(\boldsymbol{r}, E = eV) \equiv g(\boldsymbol{r}, E)$, as a function of both tip location and electron energy with atomic resolution and register.

# SI Figure Captions

**Fig. S1.**

**A.** Hypothetical set of random magnetic dopant atom locations $r_i$ on a crystal termination layer.

**B.** RKKY interaction strength $\Phi_z(r)$ for $k_F = 0.06(\pi/a_0)$. Inset shows $\Phi_z(r)$ in logarithmic scale at $r \leq 3$ nm. Spacing between the two blue arrows represents the average nearest-neighbor distance $a_m \sim 1$ nm in Fig. S2A for an average magnetic doping concentration $n_0 \approx 0.25$ nm$^{-2}$.

**C.** Calculated surface-normal spin polarization $S_z(r)$ for S=3/2 magnetic dopant atoms distributed as in S1A.

**D.** Calculated surface-normal magnetic field $B_z(r)$ calculated from S1C for S=3/2 magnetic dopant atoms distributed as in S1A; mean spin polarization $\bar{S}_z = 1.45$.

**E.** Calculated heterogeneous surface-normal magnetization $M_z(r)$ from S1D.

**F.** Calculated heterogeneous Dirac-mass gap $\Delta(r)$ from S1E.

**Fig. S2.**

Identification of the Cr dopants. Characteristic dark triangles, due to the Cr atoms substituting Bi/Sb are individually indicated with yellow arrows.

**Fig. S3.**

**A.** Measured $K(q, E=0$ meV) typical of our $Cr_x(Bi_{0.1}Sb_{0.9})_{2-x}Te_3$ samples. Magnitude of scattering interference wavevector $q_F$ due to smallest $k_F$ is shown.

**B**. Determination of $k_F$. (A) Line profiles of $K(\mathbf{q}, E=0$ meV) taken from Fig. S3A and Mov. S1 along $\bar{\Gamma} - \bar{M}$ and $\bar{\Gamma} - \bar{K}$, where $\bar{\Gamma}$ is the center of $K(\mathbf{q}, E=0$ meV).

**C.** Azimuthally averaged $K(\mathbf{q}, E=0$ meV) (red curve) is fitted with two Gaussian functions, A (green) and B (cyan) in the lower panel, in order to determine $q_F$ which is $2w_A$ of Gaussian function A.

**Fig. S4.**

**A**. Measured $\mathbf{K}(q_x, q_y = 0, E)$ typical of our $Cr_x(Bi_{0.1}Sb_{0.9})_{2-x}Te_3$ samples. The dispersion of surface states QPI is manifest.

**B**. Same $\mathbf{K}(q_x = 0, q_y, E)$. Now white curves indicate fitted dispersion of Dirac surface states given by Eq. S7 with $v = 3.3$ eV·Å. Yellow dots indicate the expected dispersion of QPI from the ARPES measurements on samples from same batch as shown in Fig. S4C.

**C**. ARPES spectrum now showing the measured dispersion form ARPES studies of crystals from same batch showing $v = 3.3 \pm 0.1$ eV·Å.

**Fig. S5.**

**A**. $AC_\Delta(\mathbf{r})$, normalized auto-correlation of $\Delta(\mathbf{r})$ in Fig. 3C.

**B**. Azimuthally averaged $AC_\Delta(\mathbf{r})$. Lorentzian fit provides FWHM, $2w = 1.24$ nm.

**C**. Normalized cross-correlation $XC_{n:\Delta}(\xi, \zeta)$. The white cross indicates the maximum of XC.

**Fig. S6.**

**A**. Histogram of $E_D(r)$ in Fig. 3F.

**B**. $AC_{E_D}(r)$, normalized auto-correlation of $E_D(r)$ in Fig. 3F. No local correlation is observed at the center.

**C**. Normalized cross-correlation $XC_{n:E_D}(\xi)$ which has the maximum at $\xi_n = 1.37$ nm as the correlation length of Cr dopant density fluctuation associated with Dirac point disorder.

**Fig. S7.**

**A**. Out-of-plane Magnetization (*M*)-Temperature (*T*) curve measured under conditions of field cooling in an applied field of 10 Oe. $T_C \sim 18$ K is determined as the turning point of the *M-T* curve.

**B**. Out-of-plane magnetic hysteresis loop measured at 4.5 K.

## SI Movie Captions

**Mov. S1.** Differential conductance map $g(r, E)$ and its associated $K(q, E)$ calculated by Eq. S5 are shown in the left and right frame, respectively. The energy $E$ is indicated on the top right corner of the right frame. The FOV size of $g(r, E)$ is 90 x 90 nm$^2$.

Figure S1

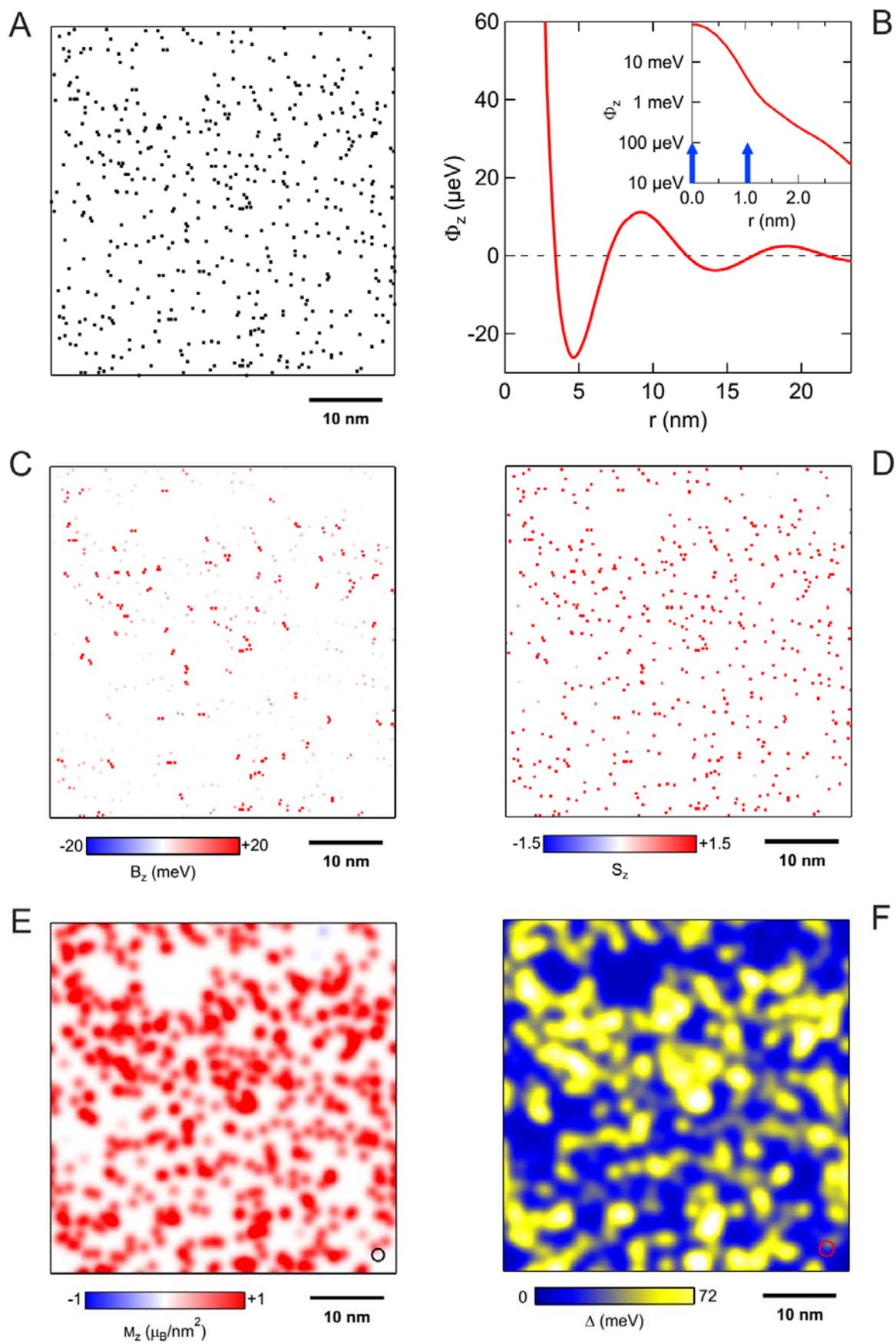

Figure S2

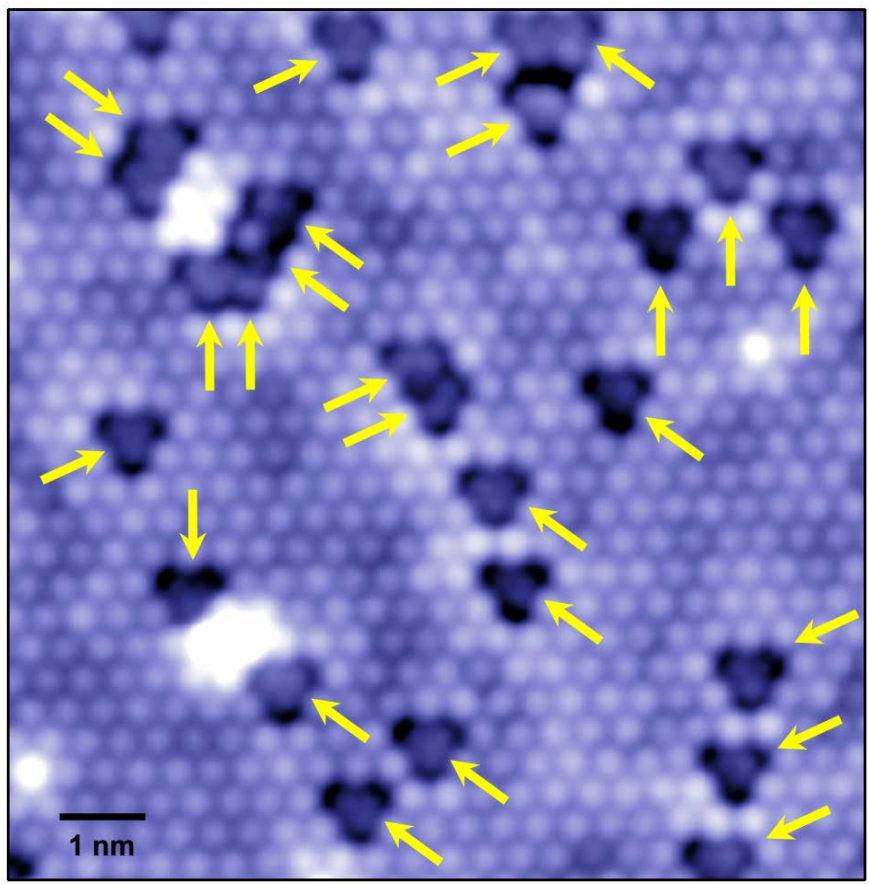

Figure S3

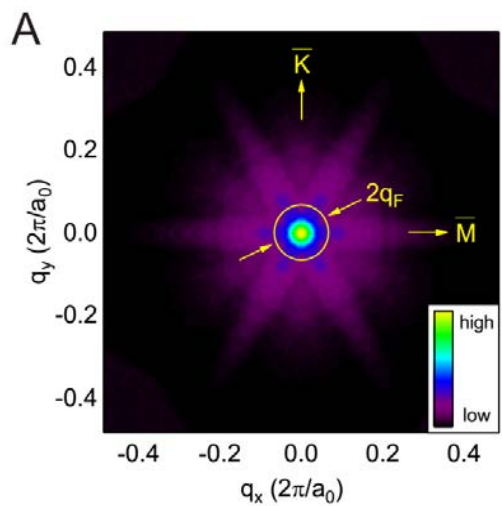

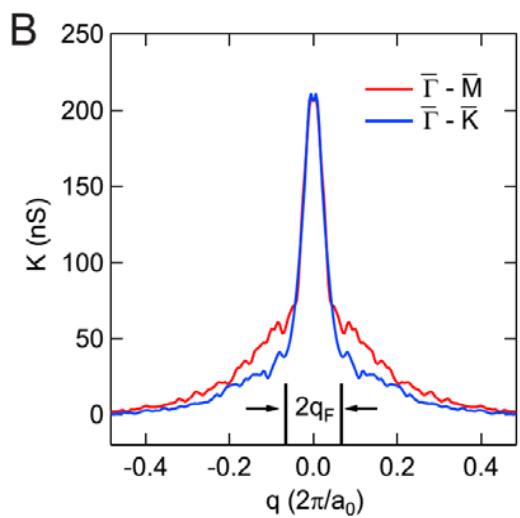
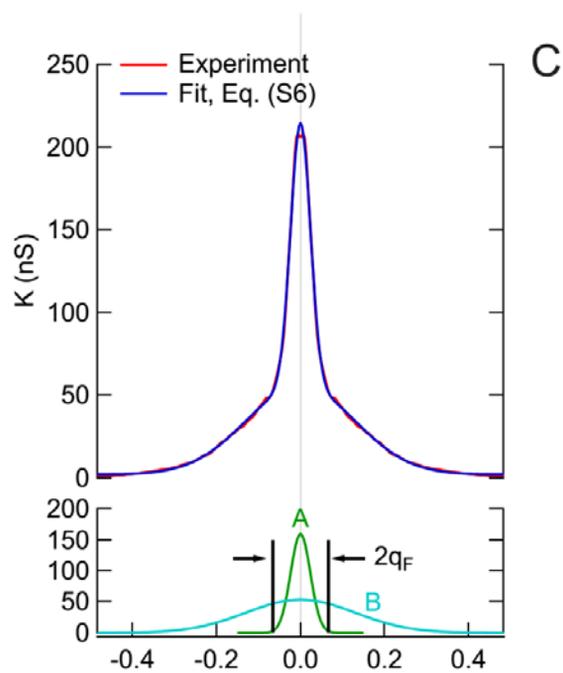

Figure S4

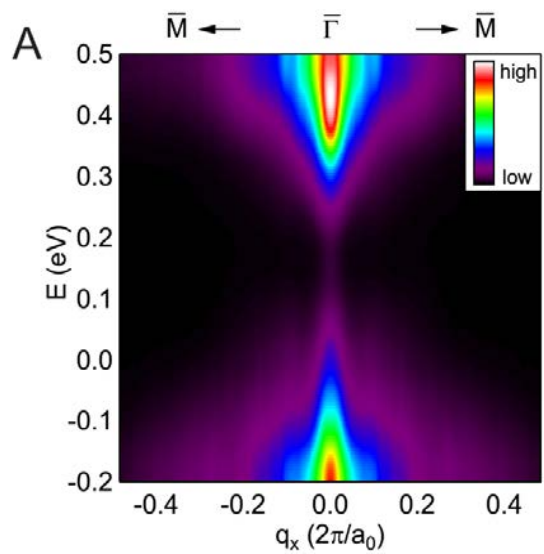

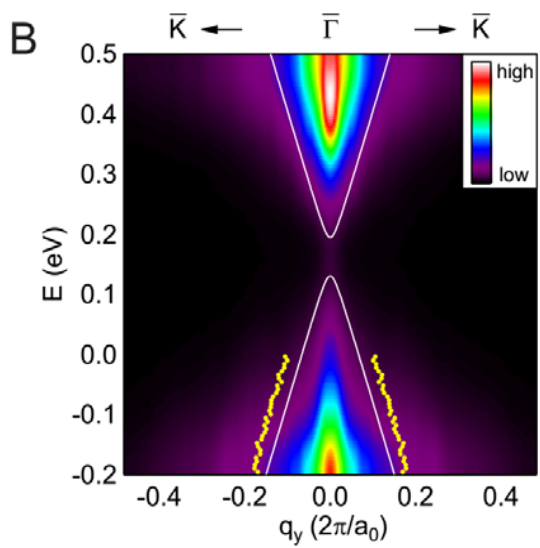

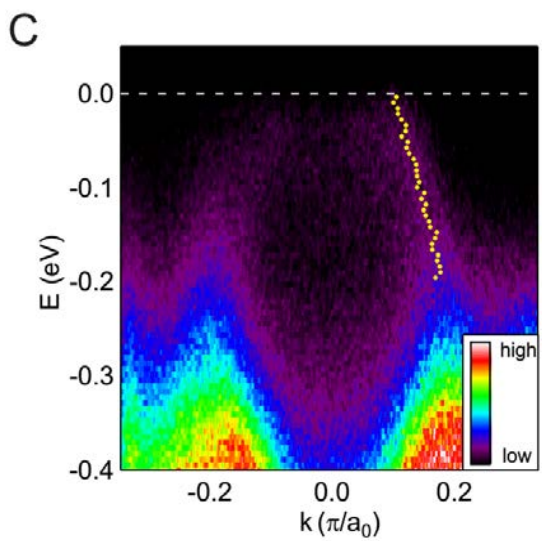

Figure S5

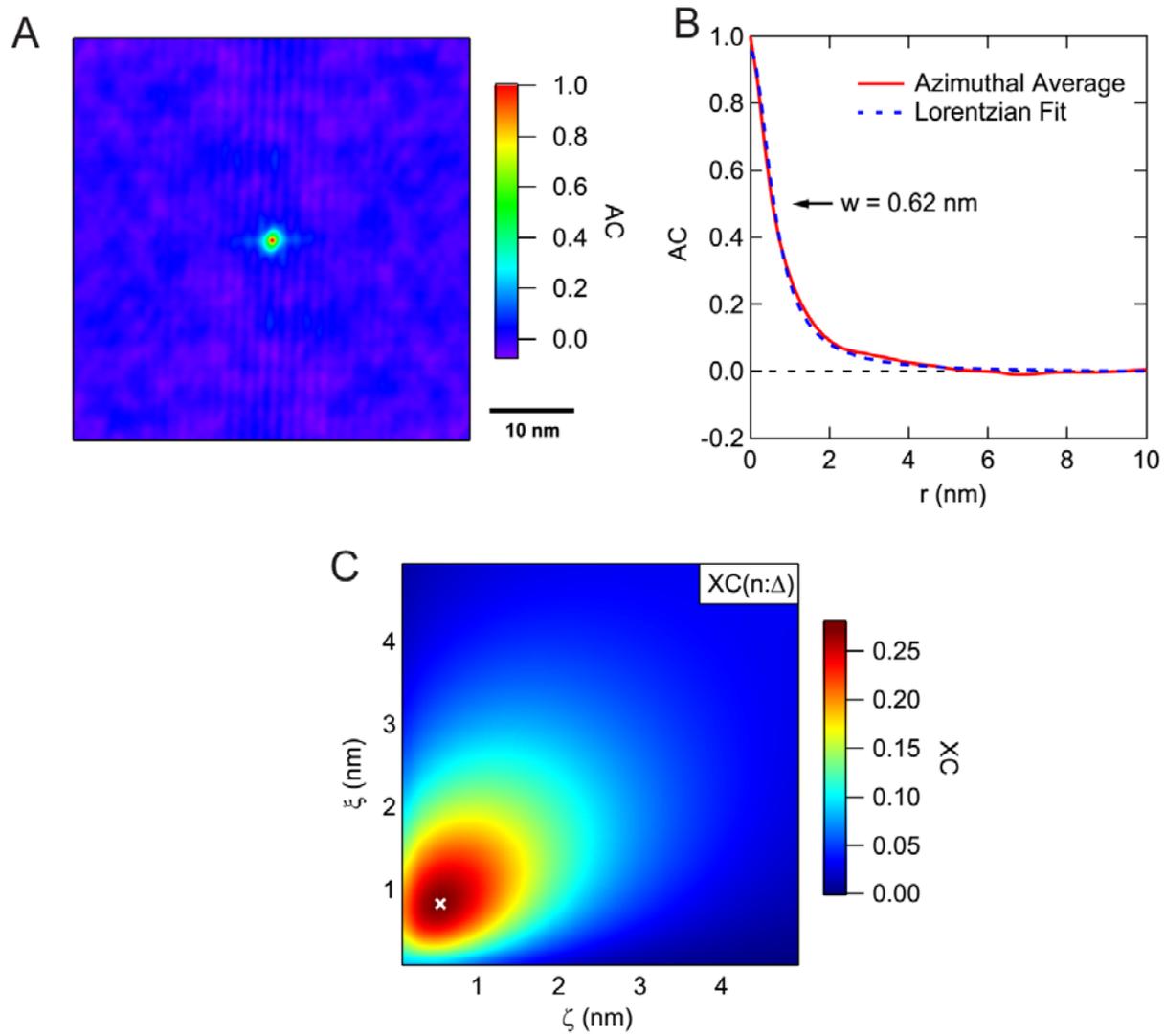

Figure S6

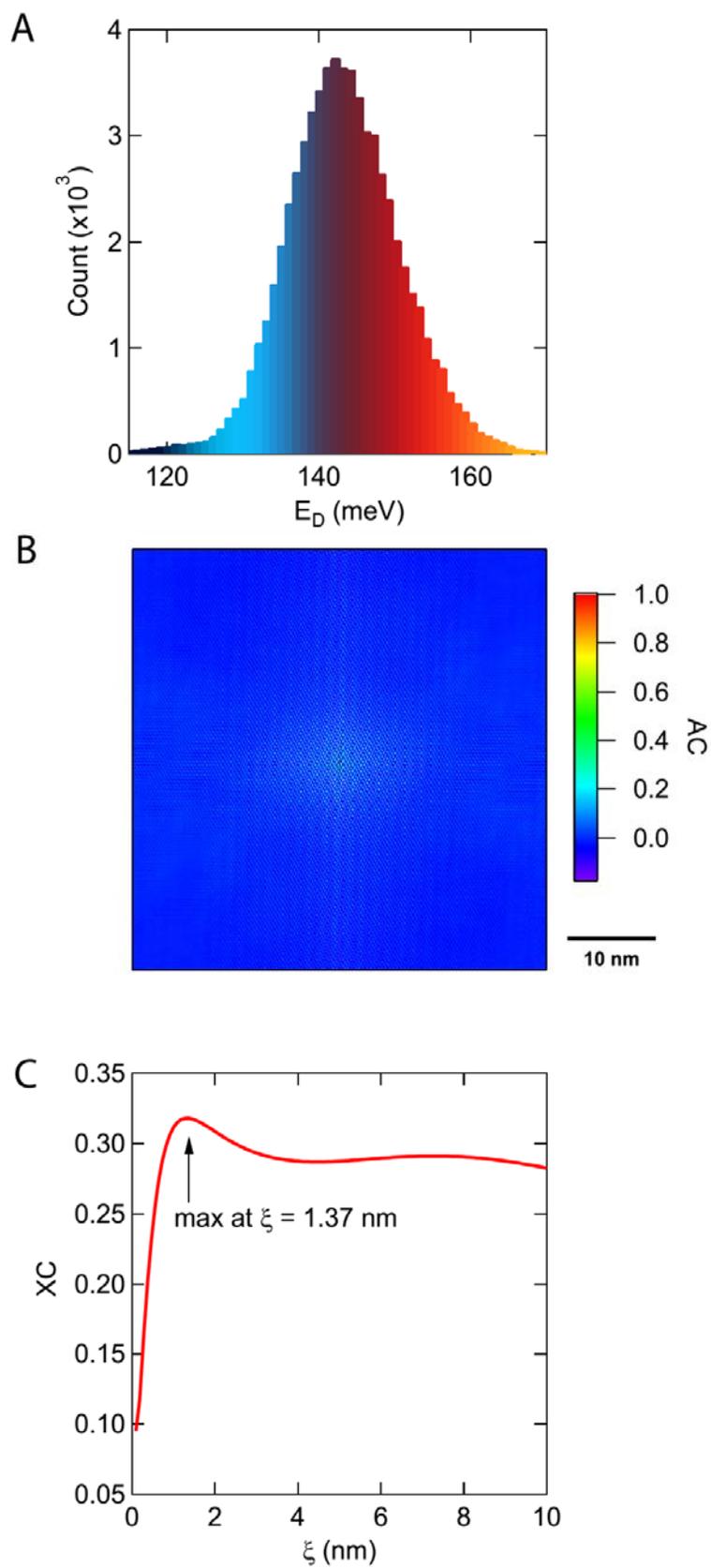

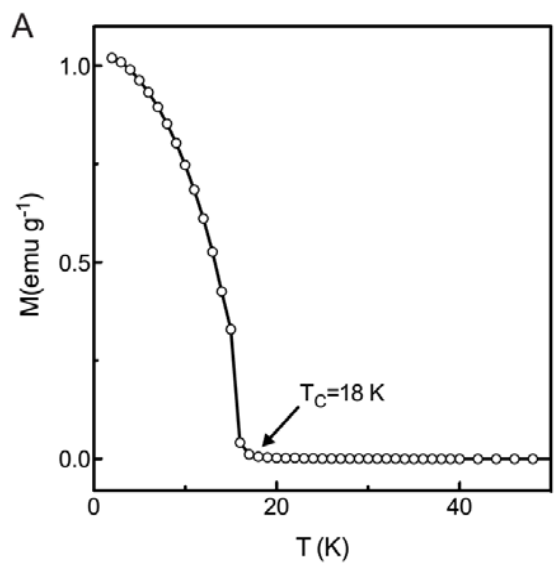

A

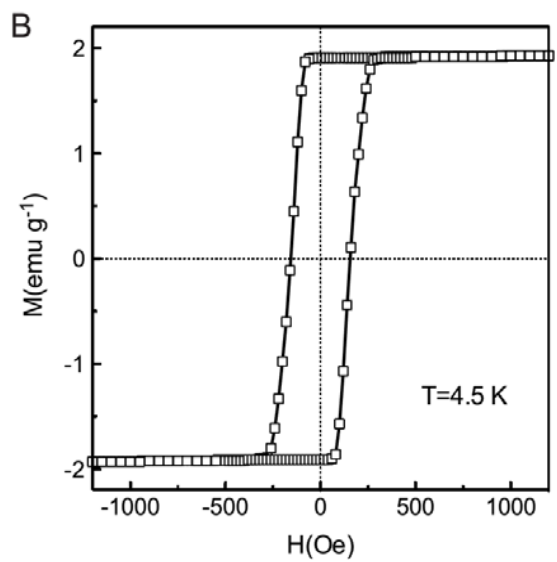

B